\documentclass[a4paper,aps,twocolumn,nofootinbib]{revtex4}
\RequirePackage[english]{babel}
\RequirePackage[latin1]{inputenc}
\RequirePackage[T1]{fontenc}
\RequirePackage{mathrsfs}
\RequirePackage{amsmath}
\RequirePackage{amssymb}
\RequirePackage{amsbsy}
\RequirePackage{bm}
\begin{document}
\title{\bf{A torsional completion of gravity for Dirac matter fields and its\\ 
applications to neutrino oscillations}}
\author{Luca Fabbri\footnote{E-mail: fabbri@dime.unige.it} 
and Stefano Vignolo\footnote{E-mail: vignolo@dime.unige.it}}
\affiliation{DIME Sez. Metodi e Modelli Matematici, Universit\`{a} di Genova,\\
Piazzale Kennedy Pad. D, 16129 Genova, Italy}
\date{\today}
\begin{abstract}
In this paper, we consider the torsional completion of gravitation for an underlying background filled with Dirac fields, applying it to the problem of neutrino oscillations: we discuss the effects of the induced torsional interactions as corrections to the neutrino oscillation mechanism.
\end{abstract}
\maketitle
\section{Introduction}
The problem of neutrino oscillations stems from the fact that known thermonuclear fusion taking place in solar processes predicts the number of neutrinos reaching earth but only a considerably smaller fraction of them is observed in our detectors. Pontecorvo has been the first who theorized that such conversion may occur whenever the leptonic numbers are not conserved if neutrinos were massive and with non-degenerate masses: the mechanism relies on the fact that mass and flavour eigenstates do not coincide, and each wave-packet of a given flavour eigenstate is the superposition of all the wave-packets of mass eigenstates, so that the wave-packets may have interference, and a phase acquires a shift \cite{p/1,p/2,g-p}. Apart from this, there are mechanisms in which neutrino oscillations may occur even in the zero-mass limit if there are external interactions of specific type, as discussed by Mikheyev and Smirnov and by Wolfenstein \cite{m-s,w}. As a consequence of this one may well wonder whether this MSW mechanism for neutrinos might also occur in the material vacuum.

In the standard wisdom of the MSW mechanism neutrino oscillations are due to external interactions with surrounding matter and therefore by construction they are absent in vacuum, but if it were possible to have neutrinos with self-interactions then, even if we define the vacuum as the absence of external interactions, nevertheless there would never be a complete vacuum for any self-interacting field. And thus one may investigate an intrinsic MSW mechanism for self-interacting neutrinos.

The torsional completion of gravity is the theory that we obtain whenever we do not constrain the most general metric-compatible connection to be symmetric in the two lower indices in holonomic basis, yielding a geometry that is endowed with torsion as well as curvature, called Cartan-Riemann geometry; when this theory is applied to Dirac spinors, torsion couples to the spin density in the same way in which curvature couples to the energy density of the Dirac matter field, giving the well known Sciama-Kibble-Einstein theory. Within such a SKE theory the Dirac matter field equation can be written in a way that is formally equivalent to the Dirac matter field equation in presence of Nambu--Jona-Lasinio potentials for every fermionic field. To have a comprehensive review of all these results\footnote{From a historical perspective, it was Cartan who first recognized the role of torsion, also called Cartan tensor, and together with the Riemann curvature, they justify the name of Cartan-Riemann geometry \cite{C1,C2,C3,C4}, while Sciama and Kibble were the first who wrote the field equations coupling torsion to the spin density in the same way in which Einstein wrote the field equations coupling curvature to energy, and so we talk about Sciama-Kibble-Einstein theory \cite{S,K}; Nambu--Jona-Lasinio potentials should be Heisenberg potentials \cite{Hehl:1971qi}, but tradition has stuck and for the sake of simplicity we will adhere to it in this paper.} and a general introduction see \cite{h-h-k-n}.

So the torsional completion of gravity for a geometry hosting Dirac matter fields provides an induced spin-contact interaction for spinor fields: this gives the opportunity to have self-interacting neutrinos mixing according to an intrinsic MSW mechanism. As a matter of fact, there have been efforts to employ the SKE theory with Dirac fields to describe left-handed massless neutrino oscillations \cite{a-d-r, FabbriNO}, but these attempts were affected by a misconception that hindered research in SKED theory for long, namely the problem of the torsion constant.

The torsion constant problem is the belief that the torsional constant must be positive and small because it is to be the Newton constant, but this misunderstanding only occurs when field equations are derived from an action that is the simplest of all; when instead the most general action is considered, it is not difficult to prove that there are a total of four constants, that is the Newton constant for the gravitational field plus three additional constants corresponding to the three irreducible decompositions of the torsion tensor \cite{Fabbri:2012yg,Fabbri:2012qr}. In those papers we have studied least-order derivative dynamics for the most general geometry, but we have not focused on the Dirac matter field in order to see what happens for the most general coupling of this spinor, which is what we will do here.

We had found that of the three irreducible components of torsion only the completely antisymmetric dual of an axial vector is excited by the spin density tensor and that there is an additional term for each flavour of the spinor that can be added to the action: so of the three constants of torsion coming from geometry there will remain only one but there will be one additional constant for each spinor that contributes to the torsion-spin coupling.

After the most general action is given, we will see what are the implications for the neutrino oscillations.
\section{SKED Theory}
In this paper, we take $(1\!+\!3)$-dimensional spacetimes filled with $\frac{1}{2}$-spin spinor fields, and all these fields will be coupled in terms of least-order derivative field equations.

All definitions we will employ hereafter are given in references \cite{Fabbri:2006xq, Fabbri:2009se, Fabbri:2008rq, Fabbri:2009yc}, but for the sake of simplicity we will give all that is needed to fix our convention in the following.

In \cite{Fabbri:2011kq,Fabbri:2013gza} we argued that torsion $Q_{\rho\mu\nu}$ is completely antisymmetric and thus it is considered to be completely antisymmetric here too: this is no loss of generality because torsion is always completely antisymmetric for Dirac fields in minimal coupling since for Dirac spinors the spin is completely antisymmetric and in least-order derivative field equations their coupling is algebraic.

From the metric tensor we define the completely antisymmetric pseudo-tensor $\varepsilon_{\rho\mu\nu\alpha}$ with which the completely antisymmetric torsion can be written according to the expression $6Q_{\rho\mu\nu}\!=\!\varepsilon_{\rho\mu\nu\alpha}W^{\alpha}$ in terms of an axial vector called axial vector torsion: and once again such a procedure is achieved without any loss of generality.

From the most general connection we can define the curvature tensor $G^{\rho}_{\phantom{\rho}\eta\alpha\nu}$ while analogously from the Levi-Civita connection it is possible to define the torsionless curvature tensor $R^{\rho}_{\phantom{\rho}\eta\alpha\nu}$ and as the most general connection can be decomposed in terms of the Levi-Civita connection plus torsion similarly the curvature tensor can be decomposed in terms of the torsionless curvature tensor supplemented by specific torsional contributions.

The curvature tensor has a single independent contraction chosen as $G^{\rho}_{\phantom{\rho}\eta\rho\nu}\!=\!G_{\eta\nu}$ with $G_{\eta\nu}g^{\eta\nu}\!=\!G$ and also the torsionless curvature tensor has a single independent contraction given by $R^{\rho}_{\phantom{\rho}\eta\rho\nu}\!=\!R_{\eta\nu}$ with $R_{\eta\nu}g^{\eta\nu}\!=\!R$ which are known as Ricci curvature tensor and scalar and torsionless curvature tensor and scalar, respectively.

Finally, we introduce $\boldsymbol{\gamma}_{a}$ known as Clifford matrices because they verify $2\boldsymbol{\mathbb{I}}\eta_{ij}\!=\!\{\boldsymbol{\gamma}_{i},\boldsymbol{\gamma}_{j}\}$ known as Clifford algebra, and $\{\boldsymbol{\gamma}_{i},[\boldsymbol{\gamma}_{j},
\boldsymbol{\gamma}_{k}]\}\!=\!4i\varepsilon_{ijkq}\boldsymbol{\pi}\boldsymbol{\gamma}^{q}$ as the relationship implicitly giving $\boldsymbol{\pi}$ as the parity-odd matrix\footnote{This matrix was historically introduced to study five dimensional extensions and it was indicated as a gamma matrix with an index five, but such a notation has no meaning in four dimensions and therefore we prefer to adopt a matrix with no index.} used to obtain the two chiral projections of the spinor field itself.

In terms of the spinorial connection we can define the spinorial covariant derivatives $\boldsymbol{D}_{\mu}$ and from the torsionless spinorial connection it is possible to define the torsionless spinorial covariant derivative $\boldsymbol{\nabla}_{\mu}$ as usual.

Because of the fact that the torsionful curvatures and covariant derivatives can be decomposed in the corresponding torsionless curvatures and covariant derivatives plus torsional contributions and spinors can be decomposed in their two chiral projections then we can employ either the full or the decomposed form in the action.

As it is well known, the original SKED theory for a Dirac particle is built on the Lagrangian given by
\begin{eqnarray}
&L\!=\!G\!-\!\frac{i}{2}(\overline{\psi}\boldsymbol{\gamma}^{\mu}\boldsymbol{D}_{\mu}\psi
\!-\!\boldsymbol{D}_{\mu}\overline{\psi}\boldsymbol{\gamma}^{\mu}\psi)\!+\!m\overline{\psi}\psi
\end{eqnarray}
or equivalently decomposed according to
\begin{eqnarray}
&L\!=\!R\!+\!\frac{1}{24}W^{2}
\!+\!\frac{1}{8}\overline{\psi} \boldsymbol{\gamma}^{\mu}\boldsymbol{\pi}\psi W_{\mu}
\!-\!i\overline{\psi}\boldsymbol{\gamma}^{\mu}\boldsymbol{\nabla}_{\mu}\psi
\!+\!m\overline{\psi}\psi
\end{eqnarray}
whose variation with respect to torsion yields
\begin{eqnarray}
&W^{\mu}\!=\!-\frac{3}{2}\overline{\psi}\boldsymbol{\gamma}^{\mu}\boldsymbol{\pi}\psi
\end{eqnarray}
which can be plugged back into the Lagrangian giving
\begin{eqnarray}
&L\!=\!R\!-\!i\overline{\psi}\boldsymbol{\gamma}^{\mu}\boldsymbol{\nabla}_{\mu}\psi
\!-\!\frac{3}{32}\overline{\psi}\boldsymbol{\gamma}^{\mu}\boldsymbol{\pi}\psi
\overline{\psi}\boldsymbol{\gamma}_{\mu}\boldsymbol{\pi}\psi\!+\!m\overline{\psi}\psi
\end{eqnarray}
showing that the torsionally-induced spin-contact interaction has a coupling constant that is positive and at the Planck scale and therefore such non-linear potential is repulsive and small; this circumstance comes from the fact that the torsion constant is assumed to be the Newton constant, and in turn this comes from the fact that the initial Lagrangian is assumed to contain the curvature alone, namely the simplest Lagrangian: instead the most general Lagrangian must include torsion contributions.

The inclusion of torsion contributions is such that the torsionless curvature and the torsional terms are independent, and consequently two constants appear, so that even if one is assumed to be the Newton constant the other is still undetermined, and the Lagrangian is
\begin{eqnarray}
&L\!=\!R\!+\!\frac{1}{128k}W^{2}
\!+\!\frac{1}{8}\overline{\psi}\boldsymbol{\gamma}^{\mu}\boldsymbol{\pi}\psi W_{\mu}
\!-\!i\overline{\psi}\boldsymbol{\gamma}^{\mu}\boldsymbol{\nabla}_{\mu}\psi
\!+\!m\overline{\psi}\psi
\end{eqnarray}
whose variation with respect to torsion yields
\begin{eqnarray}
&W^{\mu}\!=\!-8k\overline{\psi}\boldsymbol{\gamma}^{\mu}\boldsymbol{\pi}\psi
\end{eqnarray}
plugged back into the Lagrangian to give
\begin{eqnarray}
&L\!=\!R\!-\!i\overline{\psi}\boldsymbol{\gamma}^{\mu}\boldsymbol{\nabla}_{\mu}\psi
\!-\!\frac{1}{2}k\overline{\psi}\boldsymbol{\gamma}^{\mu}\boldsymbol{\pi}\psi
\overline{\psi}\boldsymbol{\gamma}_{\mu}\boldsymbol{\pi}\psi\!+\!m\overline{\psi}\psi
\end{eqnarray}
showing that the torsionally-induced spin-contact interaction has a coupling constant that is yet to be determined indeed: in \cite{Fabbri:2012yg,Fabbri:2012qr} we have studied this Lagrangian and we have discussed how under our hypotheses this is the most general in the geometric sector and in the the material sector as well; however, if we were to make the inventory of all possible terms we would witness that there is the term $\overline{\psi}\boldsymbol{\gamma}^{\mu} \boldsymbol{\pi}\psi W_{\mu}$ that could still be added to account for the interaction of geometry and matter.

When this term is included after the decomposition of the torsionful covariant derivatives into torsionless covariant derivatives and torsional terms its effect results into shifting the constant of the torsion-spin interaction by a generic amount, and then the Lagrangian becomes
\begin{eqnarray}
&L\!=\!R\!+\!\frac{1}{128k}W^{2}
\!-\!\frac{a}{8}\overline{\psi}\boldsymbol{\gamma}^{\mu}\boldsymbol{\pi}\psi W_{\mu}
\!-\!i\overline{\psi}\boldsymbol{\gamma}^{\mu}\boldsymbol{\nabla}_{\mu}\psi
\!+\!m\overline{\psi}\psi
\end{eqnarray}
whose variation with respect to torsion gives
\begin{eqnarray}
&W^{\mu}\!=\!8ka\overline{\psi}\boldsymbol{\gamma}^{\mu}\boldsymbol{\pi}\psi
\end{eqnarray}
plugged back into the Lagrangian furnishing
\begin{eqnarray}
&L\!=\!R\!-\!i\overline{\psi}\boldsymbol{\gamma}^{\mu}\boldsymbol{\nabla}_{\mu}\psi
\!-\!\frac{1}{2}ka^{2}\overline{\psi}\boldsymbol{\gamma}^{\mu}\boldsymbol{\pi}\psi
\overline{\psi}\boldsymbol{\gamma}_{\mu}\boldsymbol{\pi}\psi\!+\!m\overline{\psi}\psi
\end{eqnarray}
showing that in the effective interaction the two constants merge into a single constant, and nothing changes with respect to the case above; this situation is due to the fact that the Lagrangian is that of a single spinor field.

In the case of two spinor fields the Lagrangian is
\begin{eqnarray}
\nonumber
&L\!=\!R\!+\!\frac{1}{128k}W^{2}
\!-\!\frac{a_{1}}{8}\overline{\psi}_{1}\boldsymbol{\gamma}^{\mu}\boldsymbol{\pi}\psi_{1}W_{\mu}
\!-\!\frac{a_{2}}{8}\overline{\psi}_{2}\boldsymbol{\gamma}^{\mu} \boldsymbol{\pi}\psi_{2}W_{\mu}-\\
&-i\overline{\psi}_{1}\boldsymbol{\gamma}^{\mu}\boldsymbol{\nabla}_{\mu}\psi_{1}
\!-\!i\overline{\psi}_{2}\boldsymbol{\gamma}^{\mu}\boldsymbol{\nabla}_{\mu}\psi_{2}
\!+\!m_{1}\overline{\psi}_{1}\psi_{1}
\!+\!m_{2}\overline{\psi}_{2}\psi_{2}
\end{eqnarray}
whose variation with respect to torsion gives
\begin{eqnarray}
&W^{\mu}\!=\!8k(a_{1}\overline{\psi}_{1}\boldsymbol{\gamma}^{\mu}\boldsymbol{\pi}\psi_{1}
\!+\!a_{2}\overline{\psi}_{2}\boldsymbol{\gamma}^{\mu} \boldsymbol{\pi}\psi_{2})
\label{t}
\end{eqnarray}
plugged back into the Lagrangian furnishing
\begin{eqnarray}
\nonumber
&L\!=\!R\!-\!i\overline{\psi}_{1}\boldsymbol{\gamma}^{\mu}\boldsymbol{\nabla}_{\mu}\psi_{1}
\!-\!i\overline{\psi}_{2}\boldsymbol{\gamma}^{\mu}\boldsymbol{\nabla}_{\mu}\psi_{2}-\\
\nonumber
&-\frac{1}{2}ka_{1}^{2}\overline{\psi}_{1}\boldsymbol{\gamma}^{\mu}\boldsymbol{\pi}\psi_{1}
\overline{\psi}_{1}\boldsymbol{\gamma}_{\mu}\boldsymbol{\pi}\psi_{1}-\\
\nonumber
&-\frac{1}{2}ka_{2}^{2}\overline{\psi}_{2}\boldsymbol{\gamma}^{\mu}\boldsymbol{\pi}\psi_{2}
\overline{\psi}_{2}\boldsymbol{\gamma}_{\mu}\boldsymbol{\pi}\psi_{2}+\\
\nonumber
&+m_{1}\overline{\psi}_{1}\psi_{1}\!+\!m_{2}\overline{\psi}_{2}\psi_{2}-\\
&-ka_{1}a_{2}\overline{\psi}_{2}\boldsymbol{\gamma}^{\mu} \boldsymbol{\pi}\psi_{2}
\overline{\psi}_{1}\boldsymbol{\gamma}_{\mu}\boldsymbol{\pi}\psi_{1}
\label{l}
\end{eqnarray}
showing that in the effective interactions the three constants merge into two independent constants \cite{Fabbri:2014naa}.

It is now quite straightforward obtaining the extension to a generic number $n$ of spinorial fields: in general there would be $n$ terms of type 
$ka_{i}^{2}\overline{\psi}_{i}\boldsymbol{\gamma}^{\mu}\boldsymbol{\pi}\psi_{i} 
\overline{\psi}_{i}\boldsymbol{\gamma}_{\mu}\boldsymbol{\pi}\psi_{i}$ as self-interactions of the spinor field and $\frac{1}{2}n(n-1)$ terms of type $2ka_{i}a_{j}\overline{\psi}_{i}\boldsymbol{\gamma}^{\mu} \boldsymbol{\pi}\psi_{i}\overline{\psi}_{j}\boldsymbol{\gamma}_{\mu} \boldsymbol{\pi}\psi_{j}$ with $i\!\neq\!j$ as symmetric mutual interactions between different spinor fields.
\section{Neutrino Oscillations}
In this section we employ (\ref{l}) neglecting gravity.

Including the relative phases that arise from the fact that flavour basis and mass basis do not coincide we may focus on the contributions that mix flavours obtaining that the mixing Hamiltonian for three flavours is
\begin{eqnarray}
&H\!=\!\sum_{ij}\overline{\nu}_{i}(U_{ij}
\!-\!ka_{i}a_{j}\boldsymbol{\gamma}^{\mu}\boldsymbol{\pi}\nu_{i}
\overline{\nu}_{j}\boldsymbol{\pi}\boldsymbol{\gamma}_{\mu})\nu_{j}
\label{no}
\end{eqnarray}
where the Latin indices run over the three labels associated to the three different flavours of neutrinos, and the matrix $U_{ij}\!-\!ka_{i}a_{j}\boldsymbol{\gamma}^{\mu}\boldsymbol{\pi}\nu_{i}
\overline{\nu}_{j}\boldsymbol{\pi}\boldsymbol{\gamma}_{\mu}$ is the combination of the constant matrix $U_{ij}$ describing kinematic phases that arise from the mass terms as usual plus the field-dependent matrix $ka_{i}a_{j}\boldsymbol{\gamma}^{\mu}\boldsymbol{\pi}\nu_{i}
\overline{\nu}_{j}\boldsymbol{\pi}\boldsymbol{\gamma}_{\mu}$ describing the dynamical phases that arise from the torsionally-induced non-linear potentials we have introduced in this model.

In reference \cite{FabbriNO} such a non-linear potential has been used to show that left-handed massless neutrinos have torsionally-induced spin-contact interactions that do give rise to oscillations; alas we have not linked the constants of the problem to the length because we failed to calculate it as a consequence of the non-linearity of the problem.

In \cite{a-d-r} however the authors deal with the non-linear potential by taking neutrinos dense enough as to make the torsional background homogeneous: in doing so the nearly constant torsion background $W_{\alpha}$ can be isolated in the Hamiltonian and as a consequence they obtain the formula yielding the phase difference which in our notation can be written in the following form
\begin{eqnarray}
\Delta\Phi\!\approx\!\left(\frac{\Delta m^{2}}{2E}\!+\!\frac{1}{4}|W^{0}\!-\!W^{3}|\right)L
\label{phase}
\end{eqnarray}
having assumed $W_{1}\!=\!W_{2}\!=\!0$ for convenience and obtaining a result that depends on the length; however in their final comments the authors remark that the phase difference would be negligible because torsion would be of the order of magnitude of the Planck scale.

This is true but valid only when in the coupling equation (\ref{t}) constants $k$ and $a_{i}$ are chosen equal to the unity in Planck units, that is in the simplest but not in the most general of the models. In general models as those considered here the constant $k$ and $a_{i}$ are not equal to unity and equation (\ref{t}) yields torsion with an order of magnitude that is not necessarily the Planck scale.

As it is clear, we can plug (\ref{t}) into (\ref{phase}), and then perform some simplification: from (\ref{t}) we see that the axial vector $\overline{\nu}\boldsymbol{\gamma}^{\mu}\boldsymbol{\pi}\nu$ is proportional to the torsion dual axial vector $W^{\mu}$ and in general vector $\overline{\nu}\boldsymbol{\gamma}^{\mu}\nu$ is proportional to the momentum density vector $P^{\mu}$ so that Fierz identities given by $\overline{\nu}\boldsymbol{\gamma}_{\mu}\nu
\overline{\nu}\boldsymbol{\gamma}^{\mu}\boldsymbol{\pi}\nu\!=\!0$ yield $P_{\mu}W^{\mu}\!=\!0$ and in the case where $W^{1}\!=\!W^{2}\!=\!0$ then $P^{0}W^{0}\!=\!P^{3}W^{3}$ holds; this tells that the momentum density and the torsion dual axial vector have relationships defined up to some proportionality factor, which can be fixed by considering the other Fierz identities $\overline{\nu}\boldsymbol{\gamma}_{\mu}\nu\overline{\nu}\boldsymbol{\gamma}^{\mu}\nu
\!+\!\overline{\nu}\boldsymbol{\gamma}_{\mu}\boldsymbol{\pi}\nu
\overline{\nu}\boldsymbol{\gamma}^{\mu}\boldsymbol{\pi}\nu\!=\!0$ with the final result that if we call $P_{\mu}\!=\!(E,0,0,-P)$ then we can write $mW_{\mu}\!=\!8kaN(P,0,0,-E)$ where we have indicated in terms of $\overline{\nu}\boldsymbol{\gamma}_{\mu}\nu
\overline{\nu}\boldsymbol{\gamma}^{\mu}\nu\!=\!N^{2}$ the matter density factor.

When these conditions are implemented, and the usual approximation $2E|E\!-\!P|\!=\!m^{2}$ accounted for, we get
\begin{eqnarray}
\Delta\Phi\!\approx\!\left(\Delta m^{2}\!+\!2kamN\right)\frac{L}{2E}
\end{eqnarray}
with the dependence on the ratio between length and energy as expected; the phase has the usual kinematic contribution as difference of the squared masses plus a new dynamical contribution proportional to the neutrino mass density distribution multiplied by the $ka$ constant.

Neutrino masses are small and taking their density makes them even smaller, so if $ka$ is of the order of unity the second contribution would be negligible, but here the product of the two constants is not necessarily unitary and for large values of $ka$ such contribution is relevant.

Moreover, not only do we have the possibility to increase the strength of the torsional background in order to render the torsional contribution relevant, but the presence in our model of an additional torsion-spin coupling parameter for each flavour makes these contributions adjustable for each flavour of neutrinos.
\section{Conclusion}
In this paper, we have presented the most general version of the SKED theory for neutrinos with three flavours, seeing that the torsionally-induced spin-contact interactions contribute to neutrino oscillations and that for three families there are three different constants giving three distinct oscillation phases; we have compared our results to the ones known in literature where oscillations are due only to neutrino masses: the comparison showed that there is an analogous dependence on the ratio length over energy, but while the latter contribution depends on the difference of the squared masses the former depends on the density of mass times the $ka$ constant. This constant does not necessarily need be unitary and in fact it can be very large; moreover, the presence of the $a$ parameter means that such a constant can be different for the diverse flavours of neutrinos. This implies that the torsional contributions can be relevant, and in this case they can be adaptable to each flavour of the neutrinos. 

Of course this does not mean that such torsional contributions really are relevant nor adaptable but at least this is a possibility we can exploit. This possibility comes from the fact that the model we have presented is the most general for torsion minimally coupled.

With torsion in minimal coupling, fermions have self-interactions whose presence ensures that there can never be vacuum and an intrinsic MSW mechanism takes place for which neutrino oscillations occur.

\end{document}